\documentclass[ reprint,nofootinbib,amsmath,amssymb,
onecolumn,
aps]{revtex4-2}
\usepackage{appendix}
\usepackage{graphicx}
\usepackage{dcolumn}
\usepackage[colorlinks,linkcolor=magenta,anchorcolor=cyan,citecolor=blue]{hyperref}
\usepackage{physics}
\usepackage{bm}

\def\be{\begin{equation}}
	\def\ee{\end{equation}}
\def\ba{\begin{eqnarray}}
	\def\ea{\end{eqnarray}}

              \def\.{\cdot}

\begin{document}
	\title{Tidal Love numbers and the dynamical instability of AdS bubbles}
	\author{Gerui Chen$^{1,2,6}$}
	\email{t20152277@csuft.edu.cn}
	\author{Yu Tian$^{3,4}$}
	\email{ytian@ucas.ac.cn}
	\author{Hongbao Zhang$^{2,5}$}
	\email{hongbaozhang@bnu.edu.cn}

	\affiliation{$^1$ College of Electronic Information and Physics, Central South University of Forestry and Technology, Changsha 410004, China\\
	$^2$School of Physics and Astronomy, Beijing Normal University, Beijing 100875, China\\
	$^3$School of Physics, Chinese Academy of Sciences, Beijing 100190, China\\
	$^4$Institute of Theoretical Physics, Chinese Academy of Sciences, Beijing 100190, China\\
	$^5$Key Laboratory of Multiscale Spin Physics, Ministry of Education, Beijing Normal University, Beijing 100875, China\\
	$^6$Hunan Province Key Laboratory of Materials Surface/Interface Science and Technology, Central South University of Forestry and Technology, Changsha 410004, China}

	\begin{abstract}
		In this work, we study non-radial perturbations of AdS bubbles and their tidal Love numbers (TLNs). The odd- and even-parity TLNs are computed up to $l=6$ in the limit $k \to \infty$. The odd-parity TLNs are found to be negative, while the even-parity TLNs are positive for $\upsilon^2_s=-1$.
		As $l$ increases, the tidal Love numbers approach zero.
		The TLNs of the even-parity sector up to order $l=41$ are also calculated over the entire parameter space of $k$, from $0$ to $\infty$. We find that in the region where $p/\sigma>0$, an increasing number of TLNs become negative as $l$ increases. For $l = 41$, the highest order we have examined, the TLNs are negative everywhere except in a narrow region very close to the zero of $p/\sigma$, which agrees well with the instability criterion in the eikonal limit for self-gravitating membranes proposed by Yang {\it et al.}\ [P. R. L. {\bf 130}, 011402 (2023)].
	\end{abstract}
	\maketitle
	\section{Introduction}
	Einstein's general theory of relativity, which explains gravity as a manifestation of curved spacetime, yields several fascinating predictions, one of which is the existence of black holes~\cite{Wald,Chandrasekhar}. However, black holes as classical solutions of Einstein gravity give rise to several puzzles, such as the black hole information paradox, the connection between black hole entropy and its area, and the singularity inside black holes, all of which reveal a profound conflict between quantum mechanics and general relativity~\cite{Hawking1}. Since then, many efforts have focused on interpreting and reconciling such puzzling aspects of black holes. Given the profound difficulties in formulating a consistent quantum version of a black hole, it is valuable to investigate alternative scenarios that do not involve traditional black holes at all. This has resulted in proposals for objects like fuzzballs~\cite{Lunin}, gravastars~\cite{Mazur,Visser}, boson stars~\cite{Jetzer1,Jetzer2,Jetzer3}, Dirac stars~\cite{Finster}, Proca stars~\cite{Brito}, fermion soliton stars~\cite{Lee,Grosso,Grosso1} and other alternatives, collectively referred to as exotic compact objects (ECOs).

	The forthcoming next-generation ground-based gravitational-wave (GW)
	detectors~\cite{Kalogera}, such as the Einstein Telescope~\cite{Maggiore,Branchesi} and
	Cosmic Explorer~\cite{Abbott,Essick}, will significantly improve the accuracy of
	measurements of the tidal deformability~\cite{Pacilio,Forteza}. This improved accuracy could unveil new physics in the GW signals, potentially leading to the confirmation of the existence of these exotic compact objects. The deformability of a self-gravitating object immersed
	in an external tidal field is measured in terms of its tidal
	Love numbers (TLNs)~\cite{Love,Murray,Will}, which provide a robust analytical framework for understanding tidal effects. Initially devised in the context of Newtonian gravity, the concept of TLNs has since been successfully extended to the full theory of general relativity~\cite{Hinderer,Poisson,Damour}, largely motivated by the prospect of measuring the TLNs of neutron stars (NSs)
	through GW detections and, in turn, understanding the
	behavior of matter at supranuclear densities~\cite{Lattimer1,Lattimer2,Lattimer3,Lattimer4,Lattimer5,Lattimer6,Lattimer7}. Tidal effects are also used to explore more fundamental questions related to
	the nature of compact objects and the behavior of gravity
	in the strong-field regime~\cite{Vitor}.

	Recently, an alternative model for gravitational collapse, inspired by string theory, was proposed~\cite{Danielsson1,Danielsson2,Danielsson3,Danielsson4,Danielsson5,Giri}. It consists of a bubble of AdS space—with an anti-de Sitter interior and a Schwarzschild exterior—that is composed of ingredients from string theory.	Such an object is modeled as an ultra-compact thin-shell that is composed of three constituents, a brane with equation of motion (EOS) $p_{\tau}=-\rho_{\tau}$, a gas of massless particles with EOS $p_{g} =\rho_{g}/2$, and a stiff fluid with EOS $p_{s}=\rho_{s}$. These constituents are required based on physical and
	kinematic grounds. The radius of the AdS bubble is close to the would-be horizon of the analogous black hole but lies at a macroscopic distance outside it, thereby separating a nonsingular interior from an asymptotically flat exterior spacetime. By using the Israel-Lanczos-Sen junction conditions~\cite{Israel,Israel2,Eric} and with some basic assumptions about the equations of state of the string matter, the radius of a non-rotating neutral AdS bubble is uniquely determined to be the Buchdahl radius, $9M/4$~\cite{Buchdahl}. For any horizon-less alternative attempting to replace a black hole, stability is the paramount concern that must be addressed.
	A first step toward addressing the stability question of AdS bubbles was taken in Ref.~\cite{Danielsson1}, in which it was argued that several ingredients are necessary for AdS bubbles to be stable under radial perturbations. Then,
	Ref.~\cite{Danielsson3} further studied the stability of
	spherically symmetric bubbles undergoing dynamical radial perturbations and accretion of matter, and examined a two parameter family of fluxes required for stability. 
	
	It is worth noting that Ref.~\cite{HY1} discovered a criterion for the  instability of a generic relativistic membrane. This criterion shows that if the in-plane pressure and the surface density in the membrane share the same sign, there is a generic warping instability for modes with sufficiently high wave numbers. The underlying physical scenario is as follows: if the pressure is positive, a local vertical displacement generates an ``anti-restoring" force that pushes the mass element away from equilibrium. Although the gravitational attraction from surrounding mass elements tends to bring it back to equilibrium, the anti-spring force always wins in the eikonal limit, leading to a series of instabilities with high wave number. With the membrane motion properly taken into account, the following dispersion relation is given approximately $\omega^2\approx -p/\sigma l^2$, which leads to exponential mode growth if $p/\sigma>0$. By applying the results to commonly studied compact objects, the authors find that a significant portion of the parameter
	space of gravastars (with a de Sitter interior and Schwarzschild exterior) and AdS bubbles is dynamically unstable, and static thin-shell wormholes always have positive pressure and
	negative surface density, which implies that they are free from
	these warping instability. Therefore, requiring that the pressure and energy density of membranes have opposite signs emerges as a powerful criterion for the stability of compact objects. The relationship between the warping instability and the negativity of tidal Love numbers is also discussed in Ref.~\cite{HY1}. A membrane with density and pressure having the same sign generically prefers configurations with higher surface curvature as they are associated with a lower energy state, if gravitational backreaction is neglected.  A membrane with an ellipsoidal shape has lower potential energy than that with a spherical shape. Mathematically, the potential energy is $ \propto Q^{2}_{ij}/(2\lambda)$, where $Q_{ij}$ is the mass quadrupole moment and $\lambda$ is the tidal Love number. Negative potential energy means that $\lambda$ is negative. Even with
	gravitational backreaction included, if it is weaker
	than the anti-spring force such that the potential energy
	is still negative, the Love number will also be negative. Thus the warping instability is connected to the negativity of tidal Love numbers, which applies to generic deformations with any $l \geq 2$. In the eikonal limit, the tidal Love number has to be negative if $p/\sigma>0$. In this paper, we investigate the  non-radial perturbations of non-rotating, neutral AdS bubbles and compute
	their higher-order TLNs over the entire parameter space of $k=-\Lambda/3$ from $0$ to $\infty$, thereby investigating their dynamical instability. Our results indicate that, in the region where $p/\sigma>0$, the TLNs tend to be negative as $l$ increases, which is consistent with the expected dynamical instability in this regime.

The plan of the paper is as follows. In Section~2, we discuss the geometry of AdS bubbles. In Section~3, the definition of tidal Love numbers in full general relativity is given. In Section~4,  we discuss gravitational perturbations and the perturbed junction conditions. In Section~5, 
higher-order TLNs are calculated over the full parameter space of $k$. A brief summary is given in Section~6.
	
	We use the geometric units $G= c= 1$.
	
\section{Spacetime geometry of AdS bubbles}
The exterior spacetime of a static, spherically symmetric, neutral AdS bubble is described by the Schwarzschild metric,
\begin{eqnarray}
	ds_{+}^2=-(1-\frac{2M}{r})dt^2+(1-\frac{2M}{r})^{-1}dr^2+r^2d\theta^2+r^2\sin^2\theta d\varphi^2,\label{121}
\end{eqnarray}
while the metric in the interior is pure AdS,
\begin{eqnarray}
	ds_{-}^{2}=-(1+k \rho^2)d\tilde{t}^2+(1+k\rho^2)^{-1}d \rho^2+\rho^2d\theta^2+\rho^2\sin^2 \theta d \varphi^2.\label{ads1}
\end{eqnarray}
Here, $(t,r,\theta,\varphi)$ are the coordinates outside the shell, and $(\tilde{t},\rho,\theta,\varphi)$ are the coordinates in the
interior of the bubble. $M$ is the ADM mass, and $k$ is related to the negative cosmological constant $\Lambda$ through {\scriptsize }$k=-\frac{\Lambda}{3}>0$. The shell is positioned at $R=\frac{9}{4}M$~\cite{Danielsson1}.

To match the two spacetimes physically at the location of the shell, two junction conditions need to be{\scriptsize } imposed~\cite{Israel,Israel2,Eric}. The shell is described by parametric equations of the form
\begin{eqnarray}
	x_{\pm}^{\mu}=x_{\pm}^{\mu}(y^a),
\end{eqnarray}
where $y^a$ are the intrinsic coordinate functions of the shell denoted as $y^a=(T,\Theta,\Phi)$, and $x_{\pm}^{\mu}$ refer to the coordinate systems used outside and inside the shell, respectively. The induced metric $\gamma_{ab}$ on the shell is defined by $\gamma_{ab}=g_{\mu\nu}e^{\mu}_{a}e^{\nu}_{b}$, where $e^\mu_a=\frac{\partial x^{\mu}}{\partial y^a}$ is a set of three independent tangent vectors to the shell. The unit normal vector to the shell, $n^{\mu}$, is another important quantity characterizing the shell and satisfies the following orthonormal relations:
\begin{eqnarray}
	n_{\mu}e^{\mu}_{a}=0, \;\; n^{\mu}n_{\mu}=1. \label{ne}
\end{eqnarray}

The first junction condition ensures that the induced metric $\gamma_{ab}$ is continuous across the shell, namely, $[[\gamma_{ab}]]=0$, where $[[\cdots]]$ denotes the jump of a quantity across the shell. The metric on the shell $\Sigma$ induced from the exterior region is
\begin{eqnarray}
	ds_{\Sigma_{+}}^2=-(1-\frac{2M}{R})dT^2+R^2d\Theta^2+R^2\sin^2\Theta d\Phi^2.\label{91}
\end{eqnarray}
For the metric in the interior, a time-rescaling, $\tilde{t}=\sqrt{\frac{1-2M/R}{1+k R^2}} t$, is needed because the proper time of the bubble must be the same for both the exterior and interior coordinate systems. Then the interior metric (\ref{ads1}) becomes
\begin{eqnarray}
	ds_{-}^{2}=-(1+k \rho^2)\frac{1-2M/R}{1+k R^2}dt^2+(1+k\rho^2)^{-1}d \rho^2+\rho^2d \theta^2+\rho^2\sin^2 \theta d\varphi^2.\label{ads2}
\end{eqnarray}
The metric induced on the shell, $\rho=R$, from the inside region is identical to the induced metric (\ref{91}), so the first junction condition, which ensures that the induced metric is continuous across the shell, is satisfied. For convenience in the following discussion, we rewrite the static, spherically symmetric metrics in Eqs.~(\ref{121}) and (\ref{ads2}) in the following unified form,
\begin{eqnarray}
	ds_0^2=-f(r)dt^2+g(r)dr^2+r^2d\theta^2+r^2\sin^2\theta d\varphi^2,\label{q1}
\end{eqnarray}
where 
\[
\begin{cases}
	f(r) =  \left(1 - \frac{2M}{r}\right), & g(r) = f(r)^{-1},  \qquad \quad   r > R, \\
	f(r) = (1+k r^2)\frac{1-2M/R}{1+k R^2}, & g(r) = (1+kr^2)^{-1}, \quad   r < R.
\end{cases}
\]
The second junction condition determines the stress energy tensor on the thin shell in terms of the jump of the extrinsic curvature across the shell,
\begin{eqnarray}
	S_{ab}=-\frac{1}{8\pi}([[K_{ab}]]-\gamma_{ab}[[K]]),\label{j1}
\end{eqnarray}
where $K_{ab}=e^{\mu}_{a}e^{\nu}_{b}\nabla_{\mu}n_{\nu}$ denotes the extrinsic curvature, and $K=\gamma_{ab}K^{ab}$. This gives
\begin{eqnarray}
	S_{TT}&=&-\frac{1}{4\pi R}[[\frac{f}{\sqrt{g}}]],\label{923} \\
	S_{\Theta\Theta}&=&\frac{R}{8\pi}[[\frac{1}{\sqrt{g}}]]+\frac{R^2}{16\pi}[[\frac{f'}{f\sqrt{g}}]],\label{922}
\end{eqnarray}
where the prime denotes a derivative with respect to the radial coordinate.

\begin{figure}[htbp]
	\centering
	\includegraphics[width=0.45\textwidth]{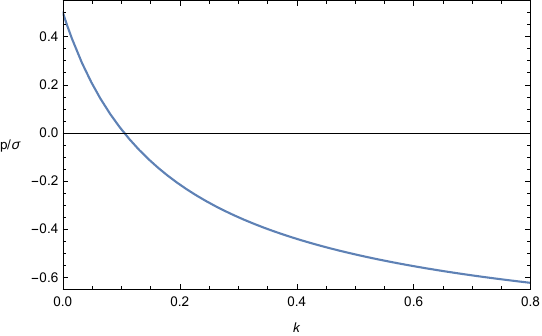}
	\caption{Ratio $p/\sigma$ as a function of the parameter $k$, with $M = 1$.}
	\label{fig1}
\end{figure}

Let us now attempt to cast $S_{ab}$ in the form of a perfect fluid. The  stress energy tensor of a perfect fluid is given by
\begin{eqnarray}
	S_{ab}=(\sigma+p)u_{a}u_{b}+p\gamma_{ab}\label{92}
\end{eqnarray}
in terms of the surface energy density $\sigma$, the surface pressure $p$, and the four velocity field $u^a$ with a non-zero component $u^T=\frac{1}{\sqrt{f(R)}}$.
By combining Eqs.~(\ref{923}), (\ref{922}) and (\ref{92}), we obtain the surface energy density and pressure:
\begin{eqnarray}
	\sigma&=&-\frac{1}{4\pi R}[[\frac{1}{\sqrt{g}}]],\\
	p&=&\frac{1}{8\pi R}[[\frac{1}{\sqrt{g}}]]+\frac{1}{16\pi}[[\frac{f'}{f\sqrt{g}}]].
\end{eqnarray}
The surface energy density and pressure of AdS bubbles, located at $R=9 M/4$, are
\begin{eqnarray}
	\sigma&=&-\frac{1}{9 \pi  M}(\frac{1}{3}-\sqrt{1+\frac{81 k M^2}{16}}),\\
	p&=&\frac{1}{108 \pi  M}(10-24 \sqrt{\frac{1}{16+81 k M^2}}-243 k M^2 \sqrt{\frac{1}{16+81 k M^2}}).
\end{eqnarray}
For a given mass of an AdS bubble, $\sigma$ is positive and increases monotonically as $k$ increases from $0$ to $\infty$. Meanwhile, $p$ decreases monotonically from positive to negative values. The ratio $p/\sigma$ as a function of $k$ is shown in Fig.~\ref{fig1}. 
It shows that $p/\sigma$ decreases monotonically from a maximum value of $0.5$ at $k = 0$; it passes through zero at $k \approx 0.1049$ and reaches a minimum value of $-1$ as $k \to \infty$.

\section{Tidal love numbers in general relativity}

In order to compute tidal Love numbers, it is necessary to derive expressions that relate the induced mass and current multipole moments to the external tidal field. This is achieved by applying linear perturbation theory to introduce a small deformation to the spacetime metric,
\begin{eqnarray}
	g_{\mu\nu}=g^{(0)}_{\mu\nu}+h_{\mu\nu},\label{full}
\end{eqnarray}
where $g^{(0)}_{\mu\nu}$ is the metric of the background spacetime (\ref{q1}), and $h_{\mu\nu}$ is a small perturbation. Owing to the spherical symmetry of the background spacetime, the perturbation $h_{\mu\nu}$ admits a decomposition into spherical harmonics and a separation into even- and odd-parity sectors, $h_{\mu\nu} = h_{\mu\nu} ^{\text{even}}+h_{\mu\nu}^{\text{odd}}$.
In the Regge-Wheeler gauge~\cite{Wheeler}, the metric perturbations are decomposed as
\begin{equation}
	h_{\mu\nu}^{\text{even}}=
	\begin{pmatrix}
		f(r)H_{0}^{lm}(t,r)Y^{lm} & H_{1}^{lm}(t,r)Y^{lm} & 0 & 0\\
		H_{1}^{lm}(t,r)Y^{lm} & g(r)H_{2}^{lm}(t,r)Y^{lm} & 0 & 0\\
		0 & 0 & r^2 K^{lm}(t,r)Y^{lm} & 0\\
		0 & 0 & 0 & r^2 \sin^2\theta K^{lm}(t,r)Y^{lm}\\
	\end{pmatrix},
\end{equation}
\begin{equation}
	h_{\mu\nu} ^{\text{odd}}=
	\begin{pmatrix}
		0 & 0 & h_{0}^{lm}(t,r)S_{\theta}^{lm} & h_{0}^{lm}(t,r)S_{\varphi}^{lm}\\
		0 & 0 & h_{1}^{lm}(t,r)S_{\theta}^{lm} & h_{1}^{lm}(t,r)S_{\varphi}^{lm}\\
		h_{0}^{lm}(t,r)S_{\theta}^{lm} & h_{1}^{lm}(t,r)S_{\theta}^{lm} & 0 & 0\\
		h_{0}^{lm}(t,r)S_{\varphi}^{lm} & h_{1}^{lm}(t,r)S_{\varphi}^{lm} & 0 & 0\\
	\end{pmatrix},
\end{equation}
with $Y^{lm}\equiv Y^{lm}(\theta,\varphi)$, $S_{\theta}^{lm}\equiv -Y_{,\varphi}^{lm}/\sin\theta$, and $S_{\varphi}^{lm}\equiv \sin\theta Y_{,\theta}^{lm}$ for notational simplicity. There is an implicit summation over the angular indices ($l,m$) here. The perturbations $H_{0}^{lm}$, $H_{1}^{lm}$, $H_{2}^{lm}$, and $K^{lm}$ belong to the even-parity sector, while $h_{0}^{lm}$ and $h_{1}^{lm}$ belong to the odd-parity sector. We consider perturbations induced by an external static tidal field. In the static background, the perturbations are consequently described by time-independent functions. Moreover, as described in Regge and Wheeler's paper~\cite{Wheeler}, there is no need to work with an arbitrary $m$, so we set $m=0$ with the advantage that $\varphi$ will  completely disappear from the calculations. Thus, we consider the following perturbations:
\begin{equation}
	h_{\mu\nu} ^{\text{even}}=
	\begin{pmatrix}
		f(r)H_{0}(r)Y(\theta) & H_{1}(r)Y(\theta) & 0 & 0\\
		H_{1}(r)Y(\theta) & g(r)H_{2}(r)Y(\theta) & 0 & 0\\
		0 & 0 & r^2 K(r)Y(\theta) & 0\\
		0 & 0 & 0 & r^2 \sin^2\theta K(r)Y(\theta)\\
	\end{pmatrix},\label{951}
\end{equation}
\begin{equation}
	h_{\mu\nu} ^{\text{odd}}=
	\begin{pmatrix}
		0 & 0 & 0 & h_{0}(r)\sin\theta Y'(\theta)\\
		0 & 0 & 0 & h_{1}(r)\sin\theta Y'(\theta)\\
		0& 0 & 0 & 0\\
		h_{0}(r)\sin\theta Y'(\theta) & h_{1}(r)\sin\theta Y'(\theta) & 0 & 0\\
	\end{pmatrix}.\label{952}
\end{equation}
Here, $Y(\theta) \equiv Y^{l0}(\theta)$, $Y'(\theta)\equiv\frac{dY(\theta)}{d\theta}$, and the superscript $l$ on the radial functions (e.g., $H_0$, $K$, etc.) is omitted for brevity. The background spacetime is spherically symmetric, so the two sectors are decoupled and can be solved independently. By solving the linearized field equations for a specific model, 
metric perturbations in (\ref{951}) and (\ref{952}) can be determined, and this problem will be discussed in the next section..

Consider an isolated,
self-gravitating compact object immersed in a tidal environment. Following Ref.~\cite{Poisson},
the symmetric and trace-free even and odd tidal multipole moments of order $l$ are defined as $\mathcal{E}_{a_{1}...a_{l}}\equiv [(l-2)!]^{-1} \langle C_{0 a_1 0 a_2 ; a_3 \dots a_l} \rangle$ and $\mathcal{B}_{a_1 \dots a_l} \equiv \left[ \frac{2}{3} (l+1) (l-2)! \right]^{-1} \langle \epsilon_{a_1 b c} C^{b c}_{a_2 0; a_3 \dots a_l} \rangle$, where $C_{abcd}$ is the Weyl tensor, a semicolon denotes a covariant derivative, $\epsilon_{abc}$ is the permutation symbol, and the angular brackets denote symmetrization of the indices $a_{i}$ and removal of all traces. The moments $\mathcal{E}_{a_{1}...a_{l}}$ (or $\mathcal{B}_{a_1 \dots a_l} $) can be decomposed in a basis of even (or odd) parity spherical harmonics. We denote by $\mathcal{E}^{lm}$ and $\mathcal{B}^{lm}$ the amplitudes of the even and odd components of the external tidal field with harmonic indices $(l,m)$, where $m$ is the azimuthal number ($|m|\leq l$).
The structure of the external tidal field is entirely encoded in the coefficients $\mathcal{E}^{lm}$ and $\mathcal{B}^{lm}$. Due to the external perturbations, 
 the mass and current multipole moments ($M_l$ and $S_l$, respectively)
of the compact object are deformed. In linear perturbation
theory, these deformations are proportional to the applied
tidal field. The remaining task is to extract the multipole moments and tidal fields from the asymptotic behavior of the full spacetime
metric given in (\ref{full}). Geroch and Hansen first provided a definition of multipole moments for axisymmetric and asymptotically flat spacetimes~\cite{Geroch,Hansen}. Then, in 1980, Thorne introduced an alternative approach to defining multipole coefficients of any spacetime metric given in asymptotically Cartesian and mass-centered coordinates~\cite{Thorne}. 
It has been shown that these two formulations of multipole moments are equivalent~\cite{YG}. The multipole moments can be extracted from the asymptotic behavior of the spacetime metric,
\begin{eqnarray}
	g_{tt}&=&-1+\frac{2M}{r}+\sum_{l=2}^{\infty} \left(\frac{2}{r^{l+1}}\left[\sqrt{\frac{4\pi}{2l+1}}M_{l}Y^{l0}+(l'<l \, \text{pole})\right]
	-\frac{2}{l(l-1)}r^{l}\left[\mathcal{E}_l Y^{l0}+(l'<l \,\text{pole})\right ]\right),\label{gtt}
\end{eqnarray}
\begin{eqnarray}
	g_{t\varphi}&=&\frac{2J}{r}\sin^2\theta+\sum_{l=2}^{\infty} \left(\frac{2}{r^l}\left[\sqrt{\frac{4\pi}{2l+1}}\frac{S_l}{l}S^{l0}_{\phi}+
	(l'<l \,\text{pole})\right]+\frac{2r^{l+1}}{3l(l-1)}\left[\mathcal{B}_l S^{l0}_{\varphi}+(l'<l \,\text{pole}) \right]\right),\label{gtb}
\end{eqnarray}
where $\mathcal{E}_l$ and $\mathcal{B}_l$ are, respectively, the amplitudes of the even and odd components of the external field with harmonic number $l$. The condition $m=0$ is set by imposing axisymmetry. 
The even-parity and odd-parity tidal Love numbers are defined as 
\begin{eqnarray}
	k_l^E&=&-\frac{1}{2}\frac{l(l-1)}{R^{2l+1}}\sqrt{\frac{4\pi}{2l+1}}\frac{M_l}{\mathcal{E}_l},\label{ke}\\
	k_l^B&=&-\frac{3}{2}\frac{l(l-1)}{(l+1)R^{2l+1}}\sqrt{\frac{4\pi}{2l+1}}\frac{S_l}{\mathcal{B}_{l}}\label{kb},
\end{eqnarray}
where $R$ is the radius of the object. This is the standard definition and is in agreement with the one used by Hinderer, Binnington and Poisson~\cite{Hinderer, Poisson}. Another commonly used definition of TLNs, proposed by Cardoso~\cite{Vitor}, is related to the standard one via
\begin{eqnarray}
	k_{l \text{C}}^{E,B}=(\frac{R}{M})^{2l+1} k_l^{E,B}.
\end{eqnarray}

\section{Gravitational perturbations and perturbed junction conditions}
 In this section, we first compute the metric perturbations of even and odd parity separately, and then discuss the perturbed junction conditions. Although the results presented here are known in the literature, we provide a detailed and pedagogical derivation to make this topic more accessible and to establish a self-contained foundation for our work.

\subsection{Gravitational perturbations}
For the even-parity perturbation, the spacetime metric is given by $g^{(0)}_{\mu\nu}+h_{\mu\nu} ^{\text{even}}$. The exterior of the AdS bubble is described by the Schwarzschild spacetime, thus $f(r)=(1-\frac{2M}{r})$ and $g(r)=(1-\frac{2M}{r})^{-1}$. The corresponding linearized Einstein equations read $\delta G_{\mu\nu}=0$.
From the combination $\delta G_{\theta\theta}-\frac{1}{\sin^2\theta}\delta G_{\phi\phi}$=0, it follows that $H_2=H_0$. For a static tidal field, the equation $\delta G_{tr}=0$ requires $H_1=0$. Using $\delta G_{r\theta}=0$, we obtain expressions for $K'$ and $K''$ in terms of $H_0$, $H_0'$, and $H_0''$. By eliminating $K$ via $\delta G_{tt}-\delta G_{rr} = 0$ and substituting the expressions for $K'$ and $K''$, we arrive at the following differential equation for $H_0 \equiv H$:
\begin{eqnarray}
	H''(r)+\frac{2 (r-M) H'(r)}{r (r-2 M)}+\frac{H(r) \left[-4 M^2+2 M r \ell
		(\ell +1)-r^2 \ell  (\ell +1)\right]}{r^2 (r-2 M)^2}=0.
\end{eqnarray}
Introducing the independent variable $x\equiv r/M-1$, the equation takes the following form~\cite{Damour}
\begin{eqnarray}
	(x^2-1)H^{''}+2xH^{'}-[\ell(\ell+1)+\frac{4}{x^2-1}]H=0,
\end{eqnarray}
where the prime denotes $d/dx$. This is an associated Legendre equation (with $\ell = m = 2$), and its solution can be written as
\begin{eqnarray}
	H(x)=c_1 \hat{P}_{\ell 2}(x) + c_2 \hat{Q}_{\ell 2}(x),\label{evenout}
\end{eqnarray}
where $c_1$ and $c_2$ are integration constants to be determined by matching with the internal solution, and $\hat{P}_{\ell 2}(x)$ and $\hat{Q}_{\ell 2}(x)$ are normalized associated Legendre functions. For convenience, we give their explicit forms as follows:
\begin{eqnarray}
	\hat{P}_{\ell 2}(x)&=&\frac{\sqrt{\pi } 2^{-\ell-1} \left(x^2-1\right) \Gamma (\ell+3) \, _2F_1\left(2-\ell,\ell+3;3;\frac{1-x}{2}\right)}{\left(4 \ell^2-8 \ell+3\right)
		\Gamma \left(\ell-\frac{3}{2}\right)},\\
	\hat{Q}_{\ell 2}(x)&=&\left(x^2-1\right) x^{-\ell-3} \, _2F_1\left(\frac{\ell+3}{2},\frac{\ell+4}{2};\ell+\frac{3}{2};\frac{1}{x^2}\right),
\end{eqnarray}
where $_2F_1 $ is the hypergeometric function. The associated Legendre functions are normalized so that
$\hat{P}_{\ell 2}\simeq x^{\ell} \simeq (r/M)^{\ell}$ and $\hat{Q}_{\ell 2}\simeq 1/x^{\ell+1} \simeq (M/r)^{\ell+1}$ when $x \to \infty$ or $r \to \infty$. Substituting $K'$ into $\delta G_{rr}=0$, we obtain an expression for $K$ in terms of $H$ and $H'$,
\begin{eqnarray}
	K(r)=\frac{2 M H'(r)}{\ell ^2+\ell -2}+\frac{H(r) \left [4 M^2+2 M r \left(\ell ^2+\ell -4\right)-r^2 \left(\ell ^2+\ell
		-2\right)\right]}{r \left(\ell ^2+\ell -2\right) (2 M-r)}.
\end{eqnarray}
The interior of the AdS bubble is described by pure AdS spacetime, with $f(r)=(1+k \rho^2)\frac{1-2M/R}{1+k R^2}$ and $g(r)=(1+k \rho^2)^{-1}$. The linearized Einstein equations read $\delta (G_{\mu\nu}+\Lambda g_{\mu\nu})=0$. From the equation $\delta (G_{\theta\theta}+\Lambda g_{\theta\theta})-\frac{1}{\sin^2\theta}\delta (G_{\phi\phi}+\Lambda g_{\phi\phi})=0$, it follows that $H_2=H_0$. Similarly, for the static tidal
field, $\delta (G_{tr}+\Lambda g_{tr})=0$ gives $H_1=0$. By using $\delta (G_{r\theta}+\Lambda g_{r\theta})=0$,  we obtain
 expressions for $K'$ and $K''$ in terms of $H_0$, $H_0'$, and $H_0''$. 
We subtract $\delta (G_{tt}+\Lambda g_{tt})=0$ from $\delta (G_{rr}+\Lambda g_{rr})=0$ to eliminate $K$. By substituting $K'$ and $K''$ into the resulting equation, we finally obtain the following differential equation for $H_0 \equiv H$,
\begin{eqnarray}
	H''(r)+\frac{\left(6-4 \Lambda  r^2\right) H'(r)}{3 r-\Lambda  r^3}+\frac{H(r) \left[2 \Lambda  r^2 \left(\Lambda  r^2-9\right)+3
		\ell ^2 \left(\Lambda  r^2-3\right)+3 \ell  \left(\Lambda  r^2-3\right)\right]}{r^2 \left(\Lambda  r^2-3\right)^2}=0.
\end{eqnarray}
This equation has two branches of solutions, and the physically meaningful solution must be regular at the origin. Thus, the inner solution is given by~\cite{Vitor},
\begin{eqnarray}
	H(r)=c_{3} \sqrt{-\Lambda} \frac{r ^{\ell } \, _2F_1\left(\frac{\ell -1}{2},\frac{\ell }{2};\ell +\frac{3}{2};\frac{\Lambda  r
			^2}{3}\right)}{3-\Lambda  r ^2},\label{ccd1}
\end{eqnarray}
where $c_3$ is a constant. By substituting $K'$ into $\delta (G_{rr}+\Lambda g_{rr})=0$, we obtain $K$  in terms of $H$ and $H'$,
\begin{eqnarray}
	K(r)=\frac{H(r ) \left[-2 \left(\Lambda ^2 r ^4+6 \Lambda  r^2-9\right)+3 \ell ^2
		\left(\Lambda  r^2-3\right)+3 \ell  \left(\Lambda  r^2-3\right)\right]-2 \Lambda
		r^3 \left(\Lambda  r^2-3\right) H'(r)}{3 \left(\ell ^2+\ell -2\right)
		\left(\Lambda  r^2-3\right)}.
\end{eqnarray}

We now turn to odd-parity perturbations, which are described by the metric $g^{(0)}_{\mu\nu}+h_{\mu\nu} ^{\text{odd}}$. The exterior of the AdS bubble is Schwarzschild spacetime. The equation $\delta G_{r\phi}=0$, with the assumption of a static tidal field, yields $h_1=0$. Application of $\delta G_{t\phi}=0$ then gives the following differential equation:
\begin{eqnarray}
	h_{0}''(r)+\frac{h_{0}(r) [r \ell  (\ell +1)-4 M]}{r^2 (2 M-r)}=0.
\end{eqnarray}
The solution was recently given in Ref.~\cite{Emanuele Berti},
\begin{eqnarray}
	h_{0}(r)&=&\tilde{c}_{1}(\frac{r}{ M})^{\ell +1}\,   _2F_1\left(-\ell +1,-\ell-2;-2\ell;\frac{2 M}{r}\right)\nonumber\\
	&+&\tilde{c}_{2}(\frac{M}{r})^\ell \,   _2F_1\left(\ell-1,\ell+2;2\ell+2;\frac{2M}{r}\right),\label{oddout}
\end{eqnarray}
where $\tilde{c}_1$ and $\tilde{c}_2$ are integration constants. As previously mentioned, the interior region is AdS spacetime. The equation $\delta (G_{r\phi}+\Lambda g_{r\phi})=0$ gives $h_1=0$. The equation $\delta (G_{t\phi}+\Lambda g_{t\phi})=0$ yields
\begin{eqnarray}
	h_{0}''(r)+\frac{h_{0}(r) \left(-2 \Lambda  r^2+3 \ell ^2+3 \ell \right)}{r^2 \left(\Lambda  r^2-3\right)}=0.
\end{eqnarray}
The solution also has two branches. In view of the regularity of the inner solution at the origin, the inner solution is given by~\cite{Vitor},
\begin{eqnarray}
	h_{0}(r)=\tilde{c}_{3}(-\Lambda)^{\frac{\ell +1}{2}}r^{\ell +1} \, _2F_1\left(\frac{\ell
	}{2}-\frac{1}{2},\frac{\ell }{2}+1;\ell
	+\frac{3}{2};\frac{\Lambda  r^2}{3}\right),\label{2337}
\end{eqnarray}
where $\tilde{c}_{3}$ is a constant.

\subsection{The junction conditions for gluing the interior and exterior  perturbed spacetimes}

To match the interior and exterior  perturbed spacetimes physically at the location of the shell, we need to impose Israel's junction conditions~\cite{Israel,Israel2,Eric}. This issue is delicate and has been addressed in several studies~\cite{Paolo1,Uchikata1,Nami,Cardoso1}.
Perturbed junction conditions are of great importance to our research on higher-order tidal Love numbers; thus, we provide a detailed review here. Our discussion follows the most recent results in Ref.~\cite{Cardoso1}.

The shell radius is perturbed by an external tidal field
\begin{eqnarray}
	\delta r_{\pm}=\delta r^{l0}_{\pm}Y^{l0}(\Theta,\Phi),
\end{eqnarray}
where we set $m=0$ in the spherical-harmonic expansion. Then the perturbed shell is located at
\begin{eqnarray}
	(T,r_0+\delta r_{\pm}Y^{l0}(\Theta,\Phi),\Theta,\Phi),
\end{eqnarray}
where $r_0$ is the location of the shell, and we omit the harmonic index $l0$ in $\delta r^{l0}_{\pm}$ from now on to avoid clutter.

We only consider linear perturbations. The $T\Phi$ component of the first junction condition $[[\gamma_{T\Phi}]]=0$ yields
\begin{eqnarray}
	[[h_0]]=0.\label{odd1}
\end{eqnarray}
Similarly, $[[\gamma_{\Theta\Theta}]]=0$ gives
\begin{eqnarray}
	\frac{2}{r_0}[[\delta r]]=-[[K]],\label{even1}
\end{eqnarray}
and $[[\gamma_{TT}]]=0$ yields
\begin{eqnarray}
	[[H_0]]=[[\frac{\delta r f^{\prime}}{f}]].\label{even2}
\end{eqnarray}

Next, we apply the second junction conditions (\ref{j1}) to the perturbation configuration. The total stress energy tensor after perturbation is
\begin{eqnarray}
	S_{ab}=(\sigma_0+\delta\sigma+p_0+\delta p)u_au_b+(p_0+\delta p)\gamma_{ab},
\end{eqnarray}
where $\sigma_0$ and $p_0$ represent unperturbed quantities, and $u_a$ and $\gamma_{ab}$ 
denote the 4-velocity and induced metric of the perturbed shell, respectively. The perturbations to the surface energy density, $\delta\sigma$, and the surface tension, $\delta p$, can be expanded as
\begin{equation}
	(\delta\sigma, \delta p)=(\delta \sigma^{lm},\delta p^{lm})Y^{lm}(\Theta,\Phi)\\
	=(\delta \sigma,\delta p)Y^{l0}(\Theta,\Phi),
\end{equation}
where we omit the harmonic index $l0$ of $\delta \sigma^{l0}$ and $\delta p^{l0}$ to simplify the notation. The tangent vectors to the perturbed shell become
\begin{equation}
	e^\mu_a=\frac{\partial x^{\mu}}{\partial y^a}=
	\begin{pmatrix}
		1 & 0 & 0 \\
		0 & \delta r Y^{l0}_{,\theta} & 0\\
		0 & 1 & 0\\
		0 & 0 & 1\\
	\end{pmatrix}.
\end{equation}
From the normalization condition $u^a u^b \gamma_{ab} = -1$ and the static condition, the four-velocity is found to be
\begin{eqnarray}
	u^a=\left(\frac{1}{\sqrt{f}}+\frac{1}{2\sqrt{f}}(H_0-\frac{f'}{f}\delta r) Y^{l0},0,0,0\right).
\end{eqnarray}
Considering the orthonormal relationship (\ref{ne}), we obtain the unit normal vector of the perturbed shell
\begin{eqnarray}
	n_a=\left(0,\sqrt{g(r)}+\frac{1}{2} \sqrt{g(r)} H_2(r) Y^{l0},-\sqrt{g}\delta r Y^{l0}_{,\theta},0\right).
\end{eqnarray}

From the $TT$ component of the second junction condition, the term with $Y^{l0}_{, \theta}$ yields
\begin{eqnarray}
	[[\sqrt{g}\delta r]]=0,\label{even3}
\end{eqnarray}
while the term with $Y^{l0}$ gives
\begin{eqnarray}
	\frac{2}{r_0^2}[[\frac{\delta r}{\sqrt{g}}]]+\frac{2}{r_0}[[\frac{H_0}{\sqrt{g}}]]+
	\frac{1}{r_0}[[\frac{H_2}{\sqrt{g}}]]-[[\frac{K^{\prime}}{\sqrt{g}}]]+\frac{1}{r_0}[[\frac{\delta r g^{\prime}}{\sqrt{g^3}}]]\nonumber\\
	-\frac{2}{r_0}[[\frac{\delta r f^{\prime}}{f\sqrt{g}}]]=8\pi \delta \sigma + 8\pi \sigma(\frac{f^{\prime}\delta r}{f} -H_0).\label{even4}
\end{eqnarray}
In the $T\Theta$ component, the term containing $Y^{l0}_{, \theta}$ yields
\begin{eqnarray}
	[[\frac{H_1}{\sqrt{g}}]]=0.\label{even5}
\end{eqnarray}
The term containing $Y^{l0}_{, \theta}$, in the $T\Phi$ component, gives
\begin{eqnarray}
	\frac{1}{2}[[\frac{h^\prime_{0}}{\sqrt{g}}]]-\frac{2}{r_0}[[\frac{1}{\sqrt{g}}]]h_0
	-\frac{1}{2}[[\frac{f^{\prime}}{f\sqrt{g}}]]h_0=8\pi \sigma h_0.\label{odd2}
\end{eqnarray}
From the $\Theta\Theta$ component, the coefficient containing $Y^{l0}$ yields
\begin{eqnarray}\label{even6}
	&&\frac{1}{r_0^2}[[\frac{\delta r}{\sqrt{g}}]]-\frac{1}{2r_0}[[\frac{H_2}{\sqrt{g}}]]+\frac{1}{r_0}[[\frac{K}{\sqrt{g}}]]-\frac{1}{4}[[\frac{H_2f^{\prime}}{f\sqrt{g}}]]
	+\frac{1}{2}[[\frac{Kf^{\prime}}{f\sqrt{g}}]]\nonumber\\
	&&+\frac{1}{2}[[\frac{K^{\prime}}{\sqrt{g}}]]-\frac{1}{2}[[\frac{H_0^{\prime}}{\sqrt{g}}]]-\frac{1}{2 r_0}[[\frac{\delta r g^{\prime}}{\sqrt{g^3}}]]+\frac{1}{r_0}[[\frac{\delta r f^{\prime}}{f\sqrt{g}}]]+
	\frac{1}{2}[[\frac{\delta r f''}{f\sqrt{g}}]]\nonumber\\
	&&-\frac{1}{2}[[\delta r (\frac{f'}{f})^2\frac{1}{\sqrt{g}}]]
	-\frac{1}{4}[[\delta r \frac{f'}{f}\frac{g'}{g^{3/2}}]]=8\pi \delta p+8\pi p(K+2\frac{\delta r}{r_0}). 
\end{eqnarray}\footnote{In equation (72) of the reference~\cite{Cardoso1},
the coefficients of the first and third terms should both be $1$. We have corrected these typos in equation (\ref{even6}).} The $\Theta \Phi$ component gives
\begin{eqnarray}
	[[\frac{h_1}{\sqrt{G}}]]=0.\label{odd3}
\end{eqnarray}
The remaining components do not yield new results, so the above equations constitute the complete set of junction conditions.

In summary, Eqs.~(\ref{odd1}), (\ref{odd2}), and (\ref{odd3}) are the junction conditions for odd parity, and Eqs.~(\ref{even1}), (\ref{even2}), (\ref{even3}), (\ref{even4}), (\ref{even5}), and (\ref{even6}) are the junction conditions for even parity.

We now verify whether the system of equations is sufficient to determine TLNs.
In the odd-parity sector, $h_1$ vanishes for static perturbations, leaving two junction conditions for three
constants of integration coming from $h_0$ (one from inside the shell and two
from outside). Since an overall factor is irrelevant to the computation of the TLNs, we already possess sufficient information to fully determine the odd-parity TLNs. For even parity, a complete description requires the addition of an equation of state relating $\delta p$ to $\delta\sigma$
\begin{eqnarray}
	\delta p=\upsilon^2_s\delta\sigma,\; \;  \upsilon^2_s\equiv(\frac{dp}{d\sigma})|_{\sigma_0},
\end{eqnarray}
where $\upsilon_s$ is the sound speed of the fluid.
For static perturbations, $H_1$ vanishes, leaving us with five junction conditions for six constants of integration (three from $H$, two from $\delta r$, and one from
$\delta \sigma$). As a result, five constants can be determined, leaving a single free constant which does not affect the computation of the even-parity TLNs.

\section{Tidal Love numbers and dynamical instability of AdS bubbles}

In this section, we employ the theoretical framework established in the preceding sections to compute tidal Love numbers of AdS bubbles. Without loss of generality, we set $M=1$, so the radius of the AdS bubble is $r_0=R=\frac{9}{4}M=\frac{9}{4}$.  We first consider odd-parity TLNs. As $r \rightarrow \infty$, the asymptotic behavior of the exterior solution (\ref{oddout}) is
\begin{eqnarray}
	h_0(r)&\sim& \tilde{c}_1(\frac{r}{M})^{l+1}+\tilde{c}_2(\frac{M}{r})^l\nonumber\\
	&=&(\frac{r}{M})^{l+1}\tilde{c}_1( 1+M^{2l+1}\frac{\tilde{c}_2}{\tilde{c}_1}\frac{1}{r^{2l+1}}).
\end{eqnarray}
By comparing this asymptotic behavior with Eq.~(\ref{gtb}) and using the definition of the odd-parity TLN~(\ref{kb}), we obtain
\begin{eqnarray}
	k_{l}^B=-\frac{1}{2}\frac{l}{l+1}(\frac{M}{R})^{2l+1}\frac{\tilde{c}_2}{\tilde{c}_1}. 
\end{eqnarray}
 The solution depends on $k$. Taking the limit $k \to \infty$, we find, to leading order $k^B_2=-0.00306225$, $k^B_3=-0.00080723$, $k^B_4=-0.000189119$, $k^B_5=-0.0000416972$, and $k^B_6=-8.88375 \times 10^{-6}$.

For even-parity TLNs, as $r \rightarrow \infty$, the asymptotic behavior of the exterior solution (\ref{evenout}) is
\begin{eqnarray}
	H_{\text{out}}\sim c_1(\frac{r}{M})^l+c_2(\frac{M}{r})^{l+1}\nonumber\\
	=c_{1}(\frac{r}{M})^l(1+M^{2l+1}\frac{c_2}{c_1}\frac{1}{r^{2l+1}}).
\end{eqnarray}
Comparing this asymptotic behavior with Eq.~(\ref{gtt}) and applying the definition of the even-parity TLN~(\ref{ke}), we obtain
\begin{eqnarray}
	k_{l}^E=\frac{1}{2}(\frac{M}{R})^{2l+1}\frac{c_2}{c_1}.
\end{eqnarray}
Following Ref.~\cite{Giri}, we take $\delta p=-\delta \sigma$ for concreteness and simplicity, which means that the fluid sound speed squared is $\upsilon^2_s=-1$. The TLN depends on the parameter $k$. In the limit $k \to \infty$, we find, to leading order
\begin{eqnarray}
	k_2^E=0.00695579\approx 0.0070.
\end{eqnarray}
The even-parity TLN calculated in Ref.~\cite{Giri} is $\lambda \approx 0.27 M^5$. Converting this to the standard definition of the TLN through $k_2^E = \frac{3}{2}\lambda R^{-5}$, we obtain
  $k_2^E\approx 0.0070$. Thus, our result is consistent with the existing literature. The even-parity TLNs up to $l=6$ are presented as follows: $k_3^E=0.00148675$, $k_4^E=0.0003043$, $k_5^E=0.0000613097$, and $k_6^E=0.0000122678$.  Both odd-parity and even-parity tidal Love numbers approach zero as $l$ increases.

\begin{figure}[htbp]
	\centering  \includegraphics[width=0.4\textwidth]{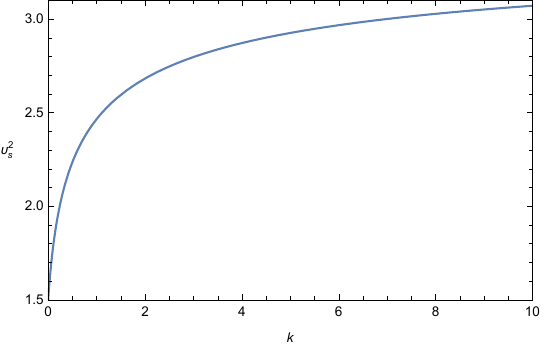}
	\caption{$\upsilon^2_s$ as a function of the parameter $k$. }
	\label{fig2}
\end{figure}

Next, we compute the even-parity TLNs up to sufficiently high order over the entire parameter space of $k$ from $0$ to $\infty$ to investigate the dynamical instability of AdS bubbles proposed in Ref.~\cite{HY1}. To obtain the even-parity tidal Love numbers for AdS bubbles with a perfect-fluid thin shell, we need to assume an equation of state for the thin-shell matter. We employ the method from Ref.~\cite{Uchikata1}.
The equilibrium sequences characterized by fixed values of $k$ are considered for AdS bubbles.
For some fixed value of $k$, the energy density and pressure of the thin shell in an equilibrium state are given by $\sigma=\sigma(R)$ and $p=p(R)$, with $R$ being the radius of the thin shell, so for each value of $k$, the corresponding equation of state is determined, $p=p(\sigma)$. The square of the speed of sound on the thin shell then is given by
\begin{eqnarray}
	\upsilon^2_s=\frac{dp}{d\sigma}=(\frac{dp}{dR}/\frac{d\sigma}{dR})|_{R=9/4}.
\end{eqnarray}
The quantity $\upsilon^2_s$ is a monotonically increasing function of $k$, as shown in Fig.~\ref{fig2}. 
The minimum value of $\upsilon^2_s$ is $1.5$ at $k=0$, and its maximum is $3.5$ in the limit $k\rightarrow\infty$. Unlike standard matter, which requires $0<\upsilon^2_s<1$, this model exhibits superluminal sound speeds.

\begin{figure}[htbp]
	\centering
	\raisebox{-.5\height}{\includegraphics[width=0.49\textwidth]{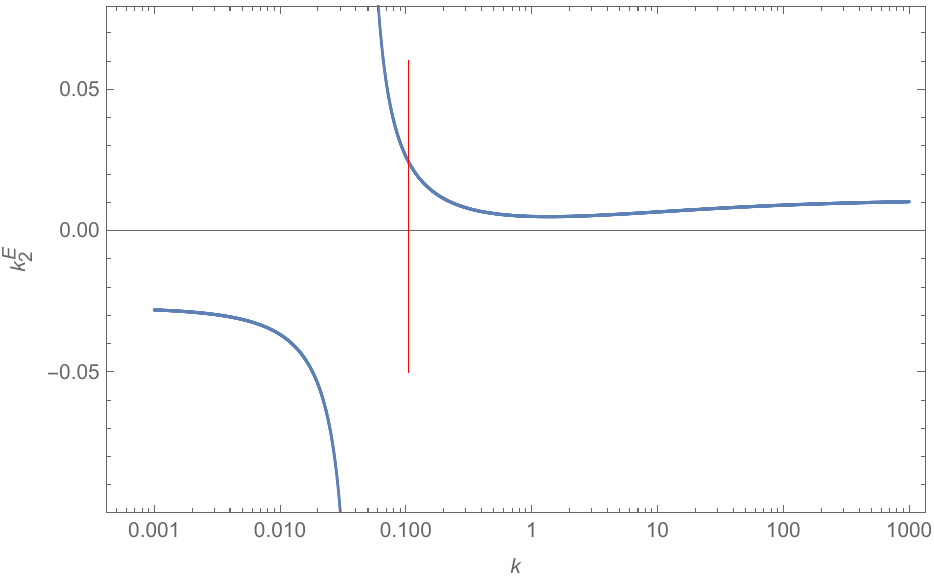}}
	\hfill
	\raisebox{-.5\height}{\includegraphics[width=0.495\textwidth]{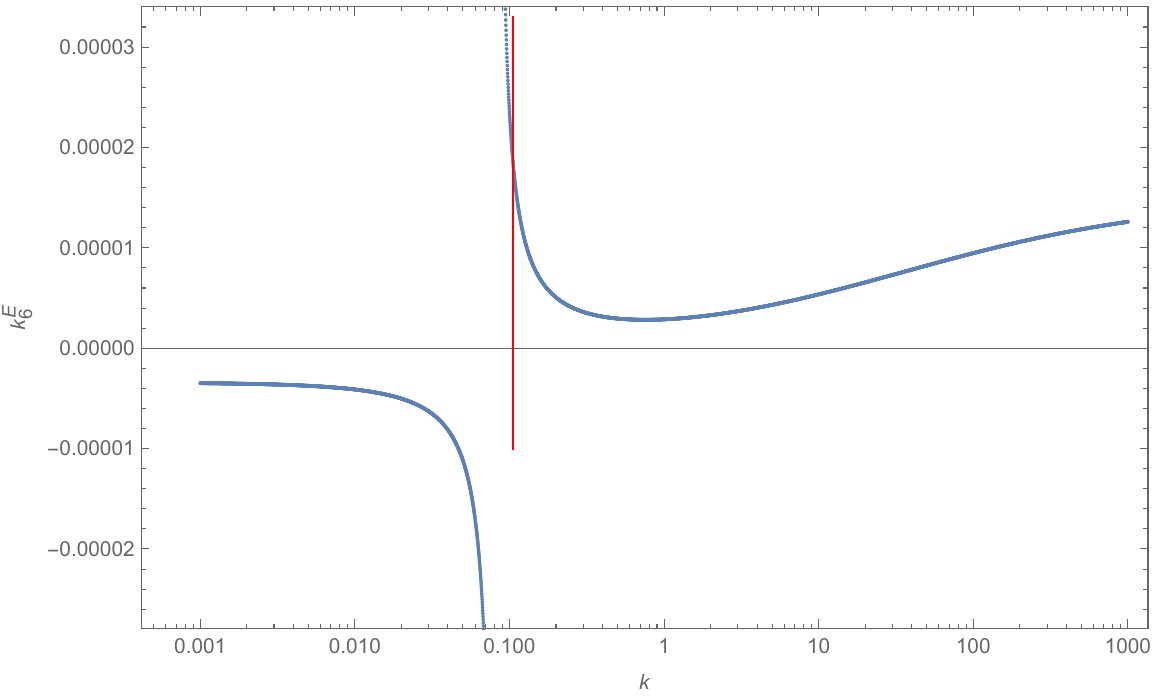}}
	\caption{ Left: Even-parity TLNs for $l=2$ over the parameter  $k$ ranging from $0$ to $\infty$. Right: Even-parity TLNs for $l=6$ over the parameter $k$ ranging from $0$ to $\infty$. The red line  marks the position where $p/\sigma=0$.}  
	\label{fig3}
\end{figure}

In principle, our framework is capable of computing tidal Love numbers to any order.
However, for very large $l$ our numerical results turn out to be unreliable, probably due to numerical errors in computing the high-order Legendre polynomials. Thus, we truncate the computation at a sufficiently high order, $l=41$. Our analysis shows that 
TLNs of different orders share a similar functional dependence on $k$. For example, TLNs for $l=2$ and $6$ are plotted explicitly in Fig.~\ref{fig3}.
The red line represents the position where  $p/\sigma=0$ at $k \approx 0.1049$.  The left region is where $p/\sigma>0$, and the right region is where $p/\sigma<0$. In the left region where $p/\sigma>0$, the TLN is negative initially. As $k$ increases, it becomes more negative until a critical point, where it discontinuously jumps to a positive value. After this jump, it decreases monotonically to a minimum value. The key distinction between the two plots is that as $l$ increases, the jump point moves closer to the zero point of $p/\sigma$. In the right region where $p/\sigma<0$, TLNs for $l=2$ and $l=6$ are positive, and the value of TLNs exhibits an initial monotonic decrease followed by an increase as $k$ becomes larger, ultimately approaching a finite limit.

\begin{figure}[htbp]
	\centering
	\includegraphics[width=0.6\textwidth]{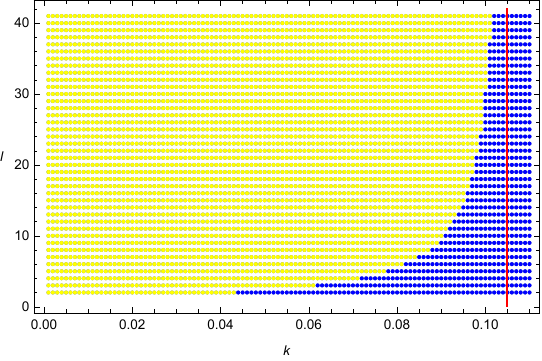}
	\caption{Even-parity TLNs up to $l = 41$ over the parameter $k$ ranging from $0$ to $0.110$, with a step size $0.001$. }
	\label{final}
\end{figure}

The tidal Love numbers of the even-parity sector up to order $l$ = 41 in the parameter space of $k$, from $0$ to $\infty$, are calculated. In the region $p/\sigma <0$ ($k>0.1409$), our calculations show that tidal Love numbers are positive for all $l$ from $2$ to $41$. This behavior is different from the static thin-shell wormholes. For static thin-shell wormholes with $p/\sigma <0$,
the TLNs for $l = 2$ and $l = 3$ are negative~\cite{Vitor}. Tidal Love numbers up to $l=41$ are shown in Fig.~\ref{final} for the parameter $k$ ranging from $0$ to $0.110$, with a step size $0.001$.
The red line also corresponds to the points where $p/\sigma = 0$. The yellow points represent negative TLNs, and the blue points represent positive TLNs. According to Ref.~\cite{HY1}, AdS bubbles in the region where $p/\sigma>0$ are unstable, corresponding to negative values of TLNs in the eikonal limit.  In the left region where $p/\sigma>0$, we find that an increasing number of TLNs become negative as $l$ increases. For $l=41$, the highest order we have examined, there are only three positive TLNs corresponding to $k=0.102$, $0.103$, and $0.104$, which are all very close to the zero point of $p/\sigma$. Thus, except for a tiny region very close to the zero point of $p/\sigma$, in the vast majority of the region where $p/\sigma>0$
(accounting for as much as $97.1\%$), our numerical results agree well with the instability criterion for self-gravitating membranes in Ref.~\cite{HY1}, namely that an instability arises in the eikonal limit, corresponding to negative values of TLNs. In practical numerical computations, as mentioned previously, $l$ cannot be very large and the maximum order $l$ we choose is $41$, so the slight discrepancy between our numerical calculations and the theoretical analysis near $p/\sigma=0$ is reasonable. Indeed, our results clearly demonstrate the tendency that the instability criterion holds when $l\to\infty$, which is just the eikonal limit used to obtain this criterion. Our study also provides more detailed and comprehensive information than the criterion
in Ref.~\cite{HY1}. As shown in Fig.~\ref{final},  the TLNs are negative even for small $l$ over a large portion of the region where $p/\sigma>0$, and an increasing number of them become negative as $l$ increases. Moreover, this behavior shifts to larger $l$ as the ratio $p/\sigma$ approaches zero.

\section{Conclusions and Summary}

AdS bubbles as a class of quantum gravity motivated black hole mimickers have attracted significant attention. For any alternative attempting to replace a black hole, stability is a critical issue. The authors of Ref.~\cite{Danielsson1} pioneered the stability analysis of AdS bubbles, and argued that several ingredients are necessary for black bubbles to be stable under radial perturbations. Recently, a criterion for the instability of a generic relativistic membrane was proposed in Ref.~\cite{HY1}. It shows that a membrane with the same sign of in-plane pressure and surface density is unstable with respect to a series of warping mode instabilities with high wave numbers. 
This warping instability is connected to the negativity of tidal Love numbers, which applies to generic deformations with any $l \geq 2$. In the eikonal limit, the tidal Love number has to be negative if $p/\sigma>0$.

In this paper, we investigate the non-radial perturbations of AdS black bubbles and compute their higher-order tidal Love numbers, thereby studying the dynamical instability proposed in Ref.~\cite{HY1}.  We first calculate
the odd- and even-parity TLNs up to $l=6$ in the limit $k \to \infty$. The even-parity TLN for $l=2$ is consistent with the existing result in Ref.~\cite{Giri}. The odd-parity TLNs are found to be negative, while the even-parity TLNs are positive for $\upsilon^2_s=-1$. As $l$ increases, the tidal Love numbers approach zero. We further compute the higher-order TLNs of even parity up to order $l=41$ over the parameter $k$ ranging from $0$ to $\infty$.
Our calculations show that tidal Love numbers of AdS bubbles are all positive from $l=2$ to $l=41$ in the region $p/\sigma <0$. In the region where $p/ \sigma > 0$, our results show that for $l = 41$, the highest order we have examined, TLNs are negative everywhere except in a narrow  region very close to the zero point of $p/\sigma$. Thus, in the region $p/ \sigma > 0$, our results agree well with the instability criterion for self-gravitating membranes in Ref.~\cite{HY1}. Our study also provides more detailed and comprehensive information than the criterion
in Ref.~\cite{HY1}. As shown in Fig.~\ref{final},  the TLNs are negative even for small $l$ over a large portion of the region where $p/\sigma>0$, and an increasing number of them become negative as $l$ increases. Moreover, this behavior shifts to larger $l$ as the ratio $p/\sigma$ approaches zero.

In summary, we present the theoretical framework to compute the tidal Love numbers of AdS bubbles in detail. We first compute the odd- and even-parity TLNs up to $l = 6$ in the limit $k \to \infty$. The odd-parity TLNs are negative, whereas the even-parity TLNs are positive for a fluid with sound speed squared $\upsilon^2_s=-1$.
As $l$ increases, the tidal Love numbers approach zero.
By computing sufficiently high-order TLNs of even parity, we find that in the region where $p/\sigma>0$, an increasing number of TLNs become negative as $l$ increases, and for $l = 41$, the highest order considered, the TLNs are negative everywhere except in a narrow region very close to the zero of $p/\sigma$. This is consistent with the instability criterion for self-gravitating membranes in Ref.~\cite{HY1}.

\begin{acknowledgments}
We thank Huan Yang for helpful discussions and the referees for their careful reading of our manuscript and constructive comments, which have significantly improved the quality of this manuscript. This work is partially supported by the National Key Research and Development
Program of China with Grant No. 2021YFC2203001 as well as the National Natural
Science Foundation of China with Grant Nos. 12361141825, 12375058, 12035016, 12475049, 11647090, and 12575047.

\end{acknowledgments}

\end{document}